\documentstyle[12pt]{article}
\begin{document}

\begin{center}
{\Large \bf Radiative Corrections for the Gauged Thirring Model in Causal Perturbation Theory}
\end{center}
\vspace{1.2cm}
\begin{center}
L. A. Manzoni and B. M. Pimentel\linebreak[1] \linebreak[1]
Instituto de F\'{\i}sica Te\'{o}rica\linebreak[1]
Universidade Estadual Paulista\linebreak[1]
Rua Pamplona, 145\linebreak[1]
01405-900 - S\~ao Paulo, S.P. \linebreak[1]
Brazil\linebreak[1]
\linebreak[1]
and
\linebreak[1]
\linebreak[1]
J. L. Tomazelli\linebreak[1] \linebreak[1]
Departamento de F\'{\i}sica e Qu\'{\i}mica - Faculdade de Engenharia\linebreak[1]
Universidade Estadual Paulista - Campus de Guaratinguet\'a\linebreak[1]
Av. Dr. Ariberto Pereira da Cunha, 333\linebreak[1]
12500-000 - Guaratinguet\'a, S.P. \linebreak[1]
Brazil
\end{center}

\vspace{1.5cm}
\begin{center}
{\bf Abstract}
\end{center}

We evaluate the one-loop fermion self-energy for the gauged Thirring model in $(2+1)$ dimensions, with one massive fermion flavor, in the framework of the causal perturbation theory. In contrast to QED$_3$, the corresponding two-point function turns out to be infrared finite on the mass shell. Then, by means of a Ward identity, we derive the on-shell vertex correction and discuss the role played by causality for nonrenormalizable theories.
\vspace{0.5cm}


keywords: Thirring model; radiative corrections; causality

\pagebreak

Recently, there has been interest in theories involving four-fermion interactions for studying dynamical fermion mass generation; such a mechanism may give a better understanding of the large top quark mass. In this context, $d$-dimensional ($2\leq d<4$) Thirring-like interactions has been considered in the large-$N$ limit\cite{gom}-\cite{sug}.

The original Thirring model\cite{thi}, a soluble model for four-fermion interaction in (1+1) dimensions, does not have any local gauge invariance. Even so,  in ref. \cite{gom} Gomes {\it et al.} have advocated a restricted gauge symmetry making use of a gauge fixing term in the Lagrangian, after linearize it by introducing an auxiliary vector field (also see \cite{hp}). Later, Itoh {\it et al.}\cite{ito} have reformulated the $d$-dimensional Thirring model as a gauge theory by introducing the hidden local symmetry (Kondo\cite{kon} has obtained the same results by using the St\"uckelberg formalism). This gauged version of the Thirring model has also been used to study fermion dynamical mass generation\cite{ito}-\cite{sug}.

In (2+1) dimensions this gauged Thirring model have a richer structure and a Chern-Simons term can be radiatively induced for odd number of fermion flavors. This parity-breaking term finds its place in many theories, from condensed matter physics to pure mathematics. In special, this term may be relevant to the fractional quantum Hall effect in the context of QED$_3$.

The gauged Thirring model in (2+1) dimensions is a nonrenormalizable theory, despite of some formal resemblance with QED$_3$. In this sense, it is not the appropriate framework to study the Hall effect. However, many questions related to the Chern-Simons term can be treated in this model. In special, we can address the regularization ambiguity related to the topological mass\cite{djt}. So, in a previous paper\cite{nos} we have considered the (2+1)-dimensional gauged Thirring model, with one massive fermion, in the framework of Epstein and Glaser's causal theory\cite{eg}\cite{sch} and studied the generation of dynamics for the gauge boson. The causal theory revealed particularly appropriate to obtaining unambiguously the coefficient of the induced Chern-Simons term.

Even more interesting is the fact that in the Epstein and Glaser's method we never run into difficulties due to ultraviolet divergencies, so that we don't have to introduce an ultraviolet cut-off. Thus, besides to offer us an unambiguous determination of the topological mass, application of the Epstein and Glaser's causal method to the gauged Thirring model in (2+1) dimensions affords an alternative approach to nonrenormalizable models.

Having this in mind, we consider the $(2+1)$-dimensional gauged Thirring model, with one massive fermion flavor, and evaluate the fermion self-energy in the framework of causal perturbation theory, calling attention for the differences with respect to the usual approaches. The fermion self-energy graph is important in the usual treatments of four-fermion interactions (such as the Nambu and Jona-Lasinio model\cite{njl}) in obtaining the gap equation. Here, we will see that the nonrenormalizability of the model will appear in a number of finite but undetermined constants.

In the sequence, we take advantage of the gauge invariance of the model to obtain the on-shell vertex correction by means of a Ward identity and verify the existence of the adiabatic limit\cite{sch}, since in a general gauge we do not have infrared divergencies.

We first set our notation and review some useful results. The original Lagrangian for the massive Thirring model\cite{thi}, considered in (2+1) dimensions, is
\begin{equation}
{\cal L} = \overline{\psi}i\gamma^{\mu}\partial_{\mu}\psi  - m\overline{\psi}\psi -\frac{G}{2}(\overline{\psi}\gamma^{\mu}\psi)(\overline{\psi}\gamma_{\mu}\psi).
\label{lorig}
\end{equation}

\noindent
We use the two dimensional realization of the Dirac algebra:
\begin{eqnarray}
\gamma^0&=&\sigma_3, \hspace{0.7cm} \gamma^1=i\sigma_1, \hspace{0.7cm} \gamma^2=i\sigma_2,\nonumber \\ 
\gamma^{\mu}\gamma^{\nu}&=&g^{\mu\nu}-i\varepsilon ^{\mu\nu\delta}\gamma_{\delta}, \hspace{1.0cm}g^{\mu\nu}={\rm diag}(1,-1,-1).
\end{eqnarray}

\noindent
where the $\sigma_j$'s are the Pauli matrices.

The theory described by the above Lagrangian does not have local gauge symmetry. However, after linearizing the interaction by introducing an auxiliary vector field, one can make use of the St\"{u}ckelberg formalism to obtain a gauge invariant version of this model. So, considering the complete BRST invariant Lagrangian in $R_{\xi}$ gauge, we have\cite{ito}\cite{kon}
\begin{equation}
{\cal L}_{Th,G}={\cal L}_{A,\psi}+{\cal L}_{\theta}+{\cal L}_{gh},
\label{ltotal}
\end{equation}

\noindent
with
\begin{eqnarray}
{\cal L}_{A,\psi} &=& \overline{\psi}i\gamma^{\mu}(\partial_{\mu}-ieA_{\mu})\psi  - m\overline{\psi}\psi +\frac{M^2}{2}A_{\mu}A^{\mu}-\frac{1}{2\xi}(\partial_{\mu}A^{\mu})^2 ,\label{lmater}\\ \nonumber \\
{\cal L}_{\theta} &=& \frac{1}{2} (\partial_{\mu}\theta)^2-\frac{\xi M^2}{2} \theta^2 ,\label{lstu} \\ \nonumber \\
{\cal L}_{gh} &=& i\left[ (\partial_{\mu}\overline{c})(\partial^{\mu}c)- \xi M^2 \overline{c}c\right],\label{lghost}
\end{eqnarray}

\noindent
where we have introduced $G=\frac{e^2}{M^2}$, with $e$ a dimensionless parameter. In these equations $A_{\mu}$ stands for the gauge boson field; $\psi$, $\overline{\psi}$ are the fermion fields and $\theta$ is the St\"{u}ckelberg neutral scalar field. Here, $c$ and $\overline{c}$ stand for the Faddeev-Popov ghost fields.

It is worth to note that the above Lagrangian was used both with a non-local and local gauge parameter in refs. \cite{ito} and \cite{kon}, respectively. In \cite{ito} the authors have considered a non-local form of the $R_{\xi}$ gauge with the purpose of analysing the fermion dynamical mass generation in the ladder approximation for the Schwinger-Dyson equation. The non-local gauge is the only one which allows for this approximation in a consistent way with the Ward-Takahashi identity for the current conservation.

However, such non-local gauges are difficult to handle and the corresponding perturbative expansion exhibit technical problems (see ref. \cite{str}). So, in order to avoid these problems, we follow here the approach of  ref. \cite{kon} and consider a local gauge parameter $\xi$. Then, it must be observed that if we take the unitary gauge (which here corresponds to making $\xi \rightarrow \infty$) the Lagrangian (\ref{lmater}) reduces to the original Thirring model, eq. (\ref{lorig}). That is, the Thirring model is just a gauge fixed version of eq. (\ref{lmater}), as pointed out in \cite{kra}.

In the limit $\xi \rightarrow \infty $, the perturbation parameter, which in the gauged model is $e$, is again $G$. But, as a consequence of the linearization, we have that graphs of the same order in $G$ are obtained as the limit of graphs of different orders in perturbation theory in $e$. For example, both the vertex correction and the box diagram (which are of order $e^3$ and $e^4$, respectively) are of  order $G^2$ when we take the limit $\xi \rightarrow \infty$.

In \cite{nos} we have proven that the singular order of an arbitrary graph in the gauged Thirring model is
\begin{equation}
\omega=3-f-\frac{3}{2}b+\frac{1}{2}n,
\label{ordemsing}
\end{equation}

\noindent
where $f$ ($b$) is the number of external fermions (bosons) and $n$ is the order of perturbation theory. From this expression we see that the Thirring model is nonrenormalizable and we expect the number of constants appearing in the general solution of the splitting problem\cite{eg}\cite{sch}, which are not fixed by causality, to increase with the order of perturbation theory. In addition, eq. (\ref{ordemsing}) gives $\omega_{se}=2$ for the singular order of the fermion self-energy and $\omega_{v}=1$ for the vertex correction.

Let us start by considering the fermion self-energy. Here we note that the tadpole contribution vanishes by charge conjugation. So, from (\ref{lmater}) we see that the first order term in the causal perturbation theory can be write down as
\begin{equation}
T_1(x)=-ie:\overline{\psi}(x)\gamma^{\mu}\psi(x):A_{\mu},
\label{acopl}
\end{equation}

\noindent
where the dots mean normal ordering of the fields in the fermion current.

From this we can construct the distribution $D_2(x_1,x_2)=R_2^{'}-A_2^{'}$ in second order of perturbation theory. So, after using Wick's theorem, the contribution for the fermion self-energy in $D_2$ is 
\begin{equation}
D_2(x_1,x_2)=:\overline{\psi}(x_1)d(y)\psi (x_2):,
\end{equation}

\noindent
where $y \equiv x_1-x_2$ and the numerical distribution
\begin{equation}
d(y)=-e^2 \gamma^{\mu}\left[ S^{(-)}(y)D_{\mu\nu}^{(+)}(-y) + S^{(+)}(y)D_{\mu\nu}^{(+)}(y) \right] \gamma^{\nu},
\label{dy}
\end{equation}

\noindent
has causal support. The fermion commutation functions in (\ref{dy}) are given by\cite{sch}

\begin{equation}
S^{(\pm )}(x)=\pm\frac{i}{(2\pi)^2}\int d^3p(p\!\!\!\slash+m)\Theta(\pm p_0)\delta (p^2-m^2)e^{-ip\cdot x},
\label{comfer}
\end{equation}

\noindent
and the boson commutation functions are\cite{nos}

\begin{equation}
D_{\mu\nu}^{(\pm )}(x)=\pm \frac{i}{(2\pi)^2}\int d^3k \frac{k_{\mu}k_{\nu}}{M^2}\delta (k^2-\xi M^2)\Theta (\pm k_0)e^{-ik\cdot x}.
\label{fcom}
\end{equation}

Now, we must perform the splitting of the causal distribution $d(y)$ into the advanced and retarded distributions $a$ and $r$, respectively. Since the splitting procedure is best done in momentum space, we go to the $p$-space and write

\begin{equation}
\hat{d}(p)= A_d(p^2) p\!\!\!\slash +B_d(p^2),
\label{dmom}
\end{equation}

\noindent
where $\hat{d}(p)$ stands for the distributional Fourier transform of $d(y)$. In this expression we have

\begin{eqnarray}
A_d(p^2)&=&\frac{1}{(2\pi )^\frac{3}{2}}\frac{e^2}{8\sqrt{p^2}M^2}{\rm sgn} (p_0) \Theta\!\!\left[p^2-a^2\right] \nonumber \\ \nonumber \\
&\times& \left[ (p^2 + m^2)\left( 1 +\frac{m^2-\xi M^2}{p^2}\right)-4m^2\right] \label{ta-},\\ \nonumber \\
B_d(p^2)&=&\frac{1}{(2\pi )^\frac{3}{2}}\frac{e^2m\xi}{4\sqrt{p^2}}{\rm sgn} (p_0) \Theta\!\!\left[p^2-a^2\right]  \label{tb-},
\end{eqnarray}

\noindent
with $a^2\equiv (m+\sqrt{\xi}M)^2$.

From these equations we can see that $\hat{d}(p)$ goes to infinite with ${\cal O}(p^2)$ when $p^2\rightarrow \infty$, that is, $\hat{d}(p)$ has singular order $\omega_{se}=2$, according to eq. (\ref{ordemsing}). Since $\omega \geq 0$, the retarded distribution is obtained by means of the ``dispersion'' formula\cite{sch}
\begin{equation}
\hat{r} (p) = \frac{i}{2\pi}\int_{- \infty}^{+ \infty} dt \frac{\hat{d}(tp)}{(t-i0)^{\omega +1}(1-t+i0)},
\label{solcentral}
\end{equation}

\noindent
which is called the central splitting (CS) solution. Then, applying this formula to (\ref{dmom}), it follows that the retarded distribution associated to fermion self-energy has the following structure
\begin{equation}
\hat{r}(p)= \tilde{A}(p^2) p\!\!\!\slash +\tilde{B}(p^2),
\label{solbip}
\end{equation}

\noindent
where 
\begin{eqnarray}
\tilde{A}(p^2)&=&\frac{i}{(2\pi )^\frac{5}{2}}\frac{e^2}{8\sqrt{p^2}M^2}\left[ \left( (p^2 + m^2)\left( 1 +\frac{m^2-\xi M^2}{p^2}\right)-4m^2\right) \right. \nonumber \\ \nonumber \\
&\times& \left.\left( \ln \left| \frac{1-\sqrt{\frac{p^2}{a^2}}}{1+\sqrt{\frac{p^2}{a^2}}}\right|- i\pi {\rm sgn}(p_0) \Theta \!\left[ p^2-a^2\right] \right) -2(2m^2+\xi M^2) \sqrt{\frac{p^2}{a^2}} \right. \nonumber \\ \nonumber \\
&+& \left. \frac{2m^2}{p^2}(m^2-\xi M^2)\left( \sqrt{\frac{p^2}{a^2}}+\frac{1}{3}\left( \frac{p^2}{a^2}\right)^{\frac{3}{2}} \right) \right] , 
\end{eqnarray}

\noindent
and
\begin{eqnarray}
\tilde{B}(p^2)\!\!\!&=&\!\!\!\frac{i}{(2\pi )^\frac{5}{2}}\frac{e^2m\xi }{4\sqrt{p^2}}\!\left[  \ln \! \left| \frac{1\!-\!\sqrt{\frac{p^2}{a^2}}}{1\!+\!\sqrt{\frac{p^2}{a^2}}}\right|- i\pi {\rm sgn}(p_0) \Theta \!\left[ p^2\!-\!a^2\right]\! +\!2\sqrt{\frac{p^2}{a^2}}\!+\! \frac{2}{3}\!\left( \frac{p^2}{a^2}\right)^{\!\!\frac{3}{2}} \right]\!\!.
\end{eqnarray}

However, for $\omega > 0$ the solution of the splitting problem is not unique. If $\tilde{r} (x)$ is the retarded part of another decomposition, then $\tilde{r}(x) - r(x)$ is a distribution with support in $\{ 0 \}$. This implies that, taking into account the minimal distribution splitting condition, the general solution for the splitting problem of a distribution with $\omega =2$, in momentum space, is given by\cite{eg}\cite{sch}:
\begin{equation}
\tilde{r}(p)=\hat{r}(p)+C_0+C_1p\!\!\!\slash + C_2p^2,
\end{equation}

\noindent
where the $C_i$'s are constants not fixed by causality.

Now, we are able to obtain the fermion self-energy, which is defined as
\begin{equation}
\hat{\Sigma}(p) = -i(2\pi )^{\frac{3}{2}}\left( \tilde{r}(p)- \hat{r}'(p)\right),
\label{self}
\end{equation}

\noindent
where $\hat{r}'(p)$ is the Fourier transform of $R'(x_1,x_2)= -e^2\gamma^{\mu}S^{(-)}(y)D_{\mu\nu}^{(+)}(-y)\gamma^{\nu}$, the first term in eq. (\ref{dy}). So that we can write
\begin{equation}
\hat{\Sigma}(p) = A(p^2)p\!\!\!\slash + B(p^2) + C_0 + C_1 p\!\!\!\slash + C_2 p^2,
\label{self1}
\end{equation}

\noindent
where we have redefined the constants to include the $-i(2\pi)^{\frac{3}{2}}$ factor. The expressions for $A(p^2)$ and $B(p^2)$ are obtained by replacing sgn($p_0$) by 1 in $\tilde{A}(p^2)$ and $\tilde{B}(p^2)$, respectively.

In contrast to the vacuum polarization tensor\cite{nos}, the Lorentz structure cannot be invoked to fix the $C_i$'s, since the splitting procedure rises the singular order of $B_d$ to that of $A_d$, for which $\omega =\omega_{se}$. Here, we can impose that the pole of the corrected fermion propagator, $S_F^{-1}(p)=(2\pi )^{\frac{3}{2}}[p\!\!\!\slash -m-(2\pi )^{-\frac{3}{2}}\hat{\Sigma}(p)]$, be in $m$. This condition, which is equivalent to $\hat{\Sigma}(p)|_{p\!\!\!\slash =m}=0$, can be satisfied choosing $C_0$ as
\begin{equation}
C_0=\frac{ie^2m^2}{(2\pi)^{\frac{5}{2}}6M^2}\left( 1+\frac{m\sqrt{\xi}M}{a^2} \right) \sqrt{\frac{m^2}{a^2}}-mC_1-m^2C_2,
\end{equation}

\noindent
so that there remains two undetermined constants.

For the moment we postpone the discussion of the remaining constants and evaluate the vertex function. But, instead to build the vertex function by considering the corresponding causal distribution in third order perturbation theory and performing its splitting in retarded and advanced distributions, we can make use of the fact that the Lagrangian (\ref{ltotal}) is BRST invariant to derive a Ward identity which will help us to obtain the on-shell three-point function, or current operator, in the limit of equal initial and final fermion momenta. 

The BRST variation for an arbitrary operator $O$ can be defined as $\delta O = \frac{\delta O}{\delta \eta}\delta \eta\equiv (\delta_{\bf B}O) \delta \eta$, where $\delta \eta$ is a Grassmann variable anticommuting with the ghost fields, $c$ and $\overline{c}$. Thus, we have that Lagrangian (\ref{ltotal}) is invariant under the set of BRST transformations
\begin{eqnarray}
{\bf {\delta}_B}\psi(x)&=&iec(x)\psi(x),\label{brpsi}\\
{\bf  {\delta}_B}\theta(x)&=&Mc(x),\label{brth} \\
{\bf  {\delta}_B}A_{\mu}(x)&=&\partial_{\mu}c(x),\label{bra} \\
{\bf  {\delta}_B}\overline{c}(x)&=&-\frac{i}{\xi}F[A,\theta ],\label{broverc}\\
{\bf  {\delta}_B}c(x)&=&0,
\end{eqnarray}

\noindent
where in the $R_{\xi}$ gauge $F[A,\theta ]=\partial_{\mu}A^{\mu}+\xi M\theta$.

We \hfill are \hfill interested \hfill in \hfill considering \hfill variations \hfill of \hfill  Green's\hfill functions \hfill of \hfill form \\ $\langle 0| T\{O(y) \overline{c}(x)\} |0\rangle $, where $O(y)$ is a shorthand notation for an arbitrary product of the fields $A_{\mu}$, $\psi$ and $\overline{\psi}$. Taking into account the invariance of Green's functions under the BRST transformation and considering $A_{\mu}$, $\psi$ and $\overline{\psi}$ as independent fields, we get
\begin{equation}
\partial_{\mu}^x\langle 0| T\{O(y) A^{\mu}(x)\} |0\rangle =i\xi\langle 0| T\{\left(\delta_{\bf B}O(y)\right) \overline{c}(x)\} |0\rangle ,
\end{equation}

\noindent
where we have made use of (\ref{broverc}) and the explicit form of $F[A,\theta ]$. Choosing $O = A^{\nu}(y)$ we obtain the following Ward identity
\begin{equation}
\partial_{\mu}^x{\cal D}^{\mu\nu}_F(x-y)=i\xi\partial_y^{\nu}\Delta^F_{gh}(y-x) ,
\label{conf1wi}
\end{equation}

\noindent
where ${\cal D}^{\mu\nu}_F$ and $\Delta^F_{gh}$ are the full Feynman propagators for the gauge boson and the ghost fields, respectively. Since the ghost fields are not interacting, $\Delta^F_{gh}$ turns out to be the free propagator. Then, taking into account the explicit form of $\Delta^F_{gh}$, we can write (\ref{conf1wi}) in momentum space as
\begin{equation}
k_{\mu}{\cal D}^{\mu\nu}_F(k) = -\xi\frac{k^{\nu}}{k^2-\xi M^2}.
\label{1wi}
\end{equation}

Following the same steps, for $O = \psi (y) \overline{\psi}(z)$, we obtain
\begin{equation}
\partial_{\mu}^x V^{\mu}(x,y,z)= e\xi S_F(y-z)\left[ \Delta^F_{gh}(y-x)-\Delta^F_{gh}(z-x)\right],
\label{2wi}
\end{equation}

\noindent
where $S_F$ is the full fermion propagator and $V^{\mu}(x,y,z)\equiv \langle 0|T\{ A^{\mu}(x)\psi(y)\overline{\psi}(z)\} |0\rangle$ is the vertex Green's function. Introducing the amputated Green's function $\Gamma^{\mu}(p', p)=\gamma^{\mu}+\Lambda^{\mu}(p', p)$, which in momentum space is related to $V^{\mu}$ by
\begin{equation}
V^{\mu}(k, p', p)=e {\cal D}^{\mu\nu}(k)S_F(p')\Gamma_{\nu}(p', p)S_F(p),
\end{equation}

\noindent
and making use of (\ref{1wi}), the Ward identity (\ref{2wi}) becomes
\begin{equation}
\Lambda^{\mu}(p, p)= -\frac{1}{(2\pi )^{\frac{3}{2}}}\frac{\partial}{\partial p_{\mu}}\hat{\Sigma} (p),
\label{difwi}
\end{equation}

\noindent
where $\Lambda^{\mu}(p',p)$ is the Fourier transform of the three-point function corresponding to the vertex correction. Thus, making use of  (\ref {difwi}) and (\ref{self}), we obtain for $\Lambda^{\mu}(p,p)$ on the mass shell 
\begin{eqnarray}
\Lambda^{\mu}(p,p)|_{p^2=m^2} &= &-\frac{C_1}{(2\pi )^\frac{3}{2}}\gamma^{\mu}-\frac{2C_2}{(2\pi )^\frac{3}{2}}p^{\mu}-\frac{e^2}{4(2\pi )^\frac{5}{2} m M^2} \left[ \gamma^{\mu}\left( -\xi M^2 \ln \left| \frac{1-\sqrt{\frac{m^2}{a^2}}}{1+\sqrt{\frac{m^2}{a^2}}}\right| \right. \right. \nonumber \\ \nonumber \\
&-& \left. \left. (2m^2+\xi M^2) \sqrt{\frac{m^2}{a^2}}+ (m^2-\xi M^2)\left(1+\frac{m^2}{3a^2}\right) \sqrt{\frac{m^2}{a^2}}\;\; \right) \right. \nonumber \\ \nonumber \\
&+& \left. p^{\mu}\left( \frac{\xi M^2}{m}\ln \left| \frac{1-\sqrt{\frac{m^2}{a^2}}}{1+\sqrt{\frac{m^2}{a^2}}}\right| +\frac{4m^2\xi M^2}{3a^3} -\frac{2(m^2-\xi M^2)}{a} \right) \right],
\label{vertice}
\end{eqnarray}

When obtaining (\ref{vertice}) through the Ward identity (\ref{difwi}), it is crucial the fact that, in a general gauge $\xi \neq 0$, the self-energy distribution is not plagued by infrared divergencies when we go on the mass shell. As a consequence, the vertex correction does not suffer from the infrared illness, which guarantees the existence of the adiabatic limit in the corresponding $S$-matrix element (we can also show that the vacuum is stable in this model\cite{nos2}).

This contrasts with QED$_3$ where, due to the massless photon, the self-energy has a logarithmic singularity on the mass shell\cite{swpt}, preventing the use of the Ward identity to obtain the on-shell vertex correction. There, one can attribute the infrared divergencies to an incorrect choice of the asymptotic states and the ill definition of the scattering operator in the asymptotic region due to the cut off from the long-range part of the interaction (see \cite{bpt} and references therein).

At this point we resume the question of the undetermined constants present in the solution for the self-energy and vertex correction distributions. We can impose a normalization condition preserving the CS solution in $p=0$ for the vertex correction, that is, $\Lambda^{\mu}(0,0)=0$, before going to the mass shell. With this condition we fix $C_1=0$, but $C_2$ still remains undetermined. This ammounts to fixing the electric form factor, leaving the fermion anomalous magnetic moment arbitrary. Here, it is important to note that the impossibility of determining all constants, by using physical considerations other than causality, is related to the nonrenormalizability of the model. So, the remaining constant $C_2$ must be regarded as a free parameter of the model.

We can also take the limit $\xi \rightarrow \infty$, in order to make the connection with the original Thirring model. This limit must be performed with $M$ finite, that is, we take $m^2 \ll \xi M^2$ and $p^2 \ll \xi M^2$. Then, we have
\begin{eqnarray}
\hat{\Sigma} (p) |_{\xi \rightarrow \infty}&\longrightarrow& C_2(p^2-m^2) , \\ \nonumber \\
\Lambda^{\mu}(p,p)|_{\xi\rightarrow \infty}&\longrightarrow& -\frac{2C_2}{(2\pi )^\frac{3}{2}}p^{\mu}.
\end{eqnarray}

\noindent
Noting that the fermion self-energy graph in the gauged Thirring model shrinks to the tadpole diagram in the original model while the vertex correction corresponds to the ``fish'' diagram with incoming momentum $q=0$, we see that in the usual treatments this limit would give divergent constants. Then, since in the causal approach we never run into divergencies, it is reasonable to expect that the result of this limit depends only on the undetermined constants.

Summing up, we have calculated the second-order correction to the current operator from the fermion self-energy in the gauged Thirring model in $(2+1)$ dimensions and showed that it is infrared safe, in contrast to QED$_3$. However, we can also show that the vacuum is stable both in the gauged Thirring model and in QED$_3$, so that we still can define free particle states\cite{nos2}.

\vspace{1.5cm}
{\bf {\large Acknowledgements}}
\vspace{0.5cm}

We would like to thank to the anonymous referee for the useful comments and calling attention for reference \cite{str}. L. A. M. is supported by Conselho Nacional de Desenvolvimento Cient\'{\i}fico e Tecnol\'{o}gico (CNPq); B. M. P. and J. L. T. are partially supported by CNPq.

\pagebreak

\end{document}